\documentclass[twocolumn,showpacs,preprintnumbers,amsmath,amssymb,superscriptaddress,aps]{revtex4}
\usepackage{graphicx}
\usepackage{dcolumn}
\usepackage{bm}
\usepackage{amsfonts}
\usepackage{mathrsfs}
\usepackage{graphicx}                   
\usepackage{dcolumn}
\usepackage{bm}
\usepackage{amsmath}

\begin{document}


\title{Nonlinear optical response from quantum kinetic equation}

\author{Zhi Li}
\email[Email:]{zhili@njust.edu.cn}
\affiliation{MIIT key Laboratory of Advanced Display Materials and Devices, Ministry of Industry and Information Technology, Institute of optoelectronics $\&$ Nanomaterials, Nanjing University of Science and Technology, Nanjing, 210094, China}
\author{Takami Tohyama}
\affiliation{Department of Applied Physics, Tokyo University of Science, Katsushika, Tokyo 125-8585, Japan}
\author{Toshiaki Iitaka}
\affiliation{Computational Engineering Applications Unit, RIKEN, 2-1 Hirosawa, Wako, Saitama 351-0198, Japan}
\author{Haibin Su}
\affiliation{Department of Chemistry, The Hong Kong University of Science and Technology, Hong Kong, China}
\author{Haibo Zeng}
\affiliation{MIIT key Laboratory of Advanced Display Materials and Devices, Ministry of Industry and Information Technology, Institute of optoelectronics $\&$ Nanomaterials, Nanjing University of Science and Technology, Nanjing, 210094, China}

\date{\today}

\begin{abstract}
 Motivated by the nonlinear Hall effect observed in topological semimetals, we studied the photocurrent by the quantum kinetic equation. We recovered the shift current and injection current discovered by Sipe et al., and the nonlinear Hall current induced by Berry curvature dipole (BCD) proposed by Inti Sodemann and Liang Fu. Especially, we further proposed that 3-form tensor like $d\vec{A}\bigwedge \vec{A}$ in which $\vec{A}$ is one-form Berry connection, can also induce photocurrent, in addition to the Berry curvature and BCD. This work will supplement the existing mechanisms for photocurrent. In contrast to the shift current induced by shift vector, all photocurrents induced by gradient/curl of Berry curvature, and high rank tensor $d\vec{A}\bigwedge \vec{A}$ require circularly polarized light and topologically non-trivial band structure, viz. non-vanishing Berry curvature.
\end{abstract}

\maketitle

\section{Introduction}
    Nonlinear optical (NLO) response, including second harmonic generation (SHG) and photogalvanic effect, has wide applications in scientific community\cite{Boyd}. For example, SHG is used for frequency doubling of laser light, and detection of the breaking of spatial inversion symmetry (SIS)\cite{Shen}, while the circular photogalvanic effect (CPGE)are used for the detection of topological charge of quantum matter\cite{CPGE1}. The dc photocurrent is resulting from frequency difference. For nonmagnetic materials, there are two different sources for photocurrent, i.e., injection current and shift current proposed by Sipe et al.\cite{NLO1,Deyo09,TM16}. The injection current is proportional to the topological charge under circularly polarized light, provided optical field with appropriate photo energy is applied\cite{JO19}. However, there is one more source for photocurrent, viz. Berry curvature dipole (BCD) proposed by Inti Sodemann and Liang Fu\cite{Fu15}. In the limit of clean system, the second order optical conductivity is inversely proportional to photon energy of optical field. Recently, Parker et al. revisited the NLO by Feynman diagram approach\cite{JEM19}. However, their formulas for NLO response are quite long, and the phyisical/geometric meaning for each contribution is still elusive.

    In periodic system, Bloch state is the function of momentum, and Berry connection $\vec{A}$ can be defined between Bloch states\cite{Xiao10}. The Berry connection plays the role of 1-form gauge field, and it can be abelian or non-abelian. Two form Berry curvature $F=d\vec{A}$ is defined by the differential of Berry connection. In materials with both SIS and time reversal invariant symmetry (TRIS), the Berry curvature is vanishing over the whole Brillouin zone (BZ). However, if SIS or TRIS is broken, Berry curvature will be non-vanishing. The band crossing in momentum space plays the role of source of Berry curvature in momentum space whose integral on closed surface surrounding the band crossing defines the topological charge\cite{AV18,BAB16}. The Weyl fermion in TaAs carries topological charge $\pm$1\cite{WHM16}, and total topological charge is zero because of TRIS\cite{Nogo}. For light-matter interaction in length gauge, electric field $\vec{E}(t)$ is couple to position matrix under Bloch states, and the off diagonal matrix element is exactly the Berry connection\cite{Resta94,Resta07}. Since NLO is related to high order of electric field $\vec{E}(t)$, and high rank tensor field will enter the resulting formulas for NLO response. Additionally, the Berry curvature dipole also cues that high rank tensor field can also induce NLO. In the view of tensor field, the geometric meaning for each contribution will be explicit. However, in velocity gauge, the advantage of tensor field may be implicit.

    Motivated by the nonlinear Hall effect induced by BCD, we studied the photocurrent by the quantum kinetic equation and classify the geometric meaning of each contribution. We recovered the shift current and injection current discovered by Sipe et al., and the nonlinear Hall current induced by BCD proposed by Inti Sodemann and Liang Fu. Especially, we further proposed that 3-form tensor like $d\vec{A}\bigwedge \vec{A}$ in which $\vec{A}$ is one-form Berry connection, can also induce photocurrent, in addition to the gradient and curl of Berry curvature. This work will supplement the existing mechanisms for photocurrent. In contrast to the shift current induced by shift vector, all photocurrents induced by gradient/curl of Berry curvature, and high rank tensor $d\vec{A}\bigwedge \vec{A}$ require circularly polarized light and topologically non-trivial band structure, viz. non-vanishing Berry curvature.

\section{quantum kinetic equation}

    Since we will make use of one-form Berry connection, length gauge will be adopted in this work. Under spatial homogeneous external field $\vec{E}(t)=\vec{E}(\omega)e^{-i\omega t}+c.c.$, the light-matter interaction can be described by below model Hamiltonian in length gauge,
    \begin{equation}\label{H}
    H(t)=\int d^{3}\vec{r}\Psi^{\dagger}(\vec{r},t)[H_{0}-e\vec{E}(t)\cdot \vec{r}]\Psi(\vec{r},t),
    \end{equation}
  where $H_{0}$ is the unperturbed Hamiltonian. The orthogonal Bloch functions $\phi_{n}(\vec{k},\vec{r})$ satisfy
  \begin{equation}\label{H0}
    H_{0}\phi_{n}(\vec{k},\vec{r})=\epsilon_{n}(\vec{k})\phi_{n}(\vec{k},\vec{r})
    \end{equation}
    in which \emph{n} is the band index and $\vec{k}$ is position in momentum space. The Bloch functions are orthogonal to each other,
    \begin{equation}\label{O}
    \int d^{3}\vec{r}\phi_{n}^{\dagger}(\vec{k},\vec{r})\phi_{m}(\vec{k'},\vec{r})=\delta_{nm}\delta(\vec{k}-\vec{k}')
    \end{equation}
   Wave function $\Psi(\vec{r},t)$ can be expressed as combination of Bloch functions with annihilation operators $a_{n}(\vec{k})$,
      \begin{equation}\label{F}
      \begin{split}
    \Psi(\vec{r},t)=\sum_{n}\int_{BZ}\frac{d^{3}\vec{k}}{(2\pi)^{3}}a_{n}(\vec{k},t)\phi_{n}(\vec{k},\vec{r})   \\
    =\sum_{n,\vec{k}}a_{n}(\vec{k},t)\phi_{n}(\vec{k},\vec{r}).
    \end{split}
    \end{equation}

   The velocity operator\cite{Xiao10} is defined as
    \begin{equation}\label{V}
    \vec{v}=\frac{i}{\hbar}[H,\vec{r}]=\frac{i}{\hbar}[H_{0},\vec{r}]-\frac{ie}{\hbar}[E_{\alpha}(t)x^{\alpha},\vec{r}],
    \end{equation}
    where we assume that summation is performed on repeated index, and the current is expressed as
    \begin{equation}\label{J}
    \begin{split}
    J^{u}(t)=\frac{ie}{\hbar}\Psi^{\dagger}(t)[H,\vec{r}]\Psi(t)=  \\
    \frac{e}{\hbar}\partial_{u}\epsilon_{n}(\vec{k})\rho_{nn}(\vec{k},t)
    +\frac{ie}{\hbar}\epsilon_{nm}(\vec{k})A_{nm}^{u}(\vec{k})\rho_{mn}(\vec{k},t) \\
    -\frac{e^{2}E_{v}(t)}{\hbar}f_{nm}^{uv}(\vec{k})\rho_{mn}(\vec{k},t)+\frac{e^{2}E_{v}(t)}{\hbar}g_{n}^{uv}(k,t)
    \end{split}
    \end{equation}
    where $\partial_{u}=\frac{\partial}{\partial k_{u}}$, energy difference $\epsilon_{nm}(\vec{k})=\epsilon_{n}(\vec{k})-\epsilon_{m}(\vec{k})$, and density matrix $\rho_{mn}(\vec{k},t)=a_{n}^{\dagger}(\vec{k},t)a_{m}(\vec{k},t)$.
    Here, we used the position matrix\cite{B1962} and the definition of Berry connection,
   \begin{equation}\label{P}
   \begin{split}
    \langle n'\vec{k}'|\vec{r}|n\vec{k}\rangle=\int d^{3}\vec{r}\phi_{n'}(\vec{k'},\vec{r})\vec{r}\phi_{n}(\vec{k},\vec{r})=  \\
    A_{n'n}(\vec{k})\delta(\vec{k}-\vec{k}')-i\delta_{n'n}\nabla_{\vec{k}}\delta(\vec{k}-\vec{k}'),
   \end{split}
    \end{equation}
    where Berry connection $\vec{A}_{nm}(\vec{k})=i\langle u_{n}(\vec{k})|\nabla_{\vec{k}}u_{m}(\vec{k})$, in which $u_{n}(\vec{k})=e^{-i\vec{k}\cdot \vec{r}}\phi_{n}(\vec{k},\vec{r})$ is the periodic part of Bloch function.

    The (static) nonabelian Berry curvature  $f_{mn}^{uv}(\vec{k})$ is defined as,
    \begin{equation}\label{NAC}
    \begin{split}
    f_{nm}^{uv}(\vec{k})=\partial_{u}A_{nm}^{v}(\vec{k})-\partial_{v}A_{nm}^{u}(\vec{k})-  \\
    i\sum_{l}[A_{nl}^{u}(\vec{k})A_{lm}^{v}(\vec{k})-A_{nl}^{v}(\vec{k})A_{lm}^{u}(\vec{k})]
    \end{split}
    \end{equation}
    which is asymmetric under the exchange of indices of $u$ and $v$, and the dynamical Berry curvature $g_{n}^{uv}(\vec{k},t)$ is defined as,
     \begin{equation}\label{Dynamical}
     \begin{split}
    g_{n}^{uv}(\vec{k},t)=i[\partial_{u}a_{n}^{\dagger}(\vec{k},t)\partial_{v}a_{n}(\vec{k},t)-  \\
    \partial_{v}a_{n}^{\dagger}(\vec{k},t)\partial_{u}a_{n}(\vec{k},t)]
    \end{split}
    \end{equation}

    The dynamics of $\rho(\vec{k},t)$ can be described by collisionless quantum kinetic equation~\cite{JR86,AHM11,ZL18,ZL19},
  \begin{equation}\label{EOM}
    -\hbar\frac{\partial\rho(\vec{k},t)}{\partial t}=e\textbf{E}\cdot\frac{\partial\rho(\vec{k},t)}{\partial \vec{k}}+i [H,\rho(\vec{k},t)].
  \end{equation}
 For intrinsic NLO effect from band structure, we ignore all the scattering term here. We expand the density matrix up to the second order of field strength,
    \begin{equation}\label{Expansion}
    \rho_{nm}(\vec{k},t)=\rho^{(0)}_{nm}(\vec{k})+\rho^{(1)}_{nm}(\vec{k},t)+\rho^{(2)}_{nm}(\vec{k},t).
    \end{equation}
  From Eq.~(10), the first-order inter-band (n$\neq$m)density matrix reads,
  \begin{equation}\label{rho1}
  \rho^{(1)}_{nm}(\vec{k},\omega)=\frac{\langle n|H_{1}(\omega)|m\rangle(\rho^{(0)}_{mm}-\rho^{(0)}_{nn})}{\hbar\omega-\epsilon_{nm}(\vec{k})},
  \end{equation}
  while the intra-band first-order density matrix reads
  \begin{equation}\label{rho1}
  \rho^{(1)}_{nn}(\vec{k},\omega)=\frac{-ie}{\hbar\omega}\vec{E}(\omega)\cdot \partial_{\vec{k}}\rho^{(0)}_{nn}(\vec{k}).
  \end{equation}
   Here, $\rho^{(0)}_{nn}=\frac{1}{1+\exp(\frac{\epsilon_{n}}{k_{B}T})}$ ($k_{B}$ Boltzmann constant, \emph{T} temperature) is fermi-Dirac distribution of band \emph{n}, and $H_{1}(\omega)=-e\vec{E}(\omega)\cdot \vec{r}$.
   From Eq.~(10), the second-order intra-band density matrix reads,
   \begin{widetext}
  \begin{equation}\label{in2nn}
   \begin{aligned}
     \rho_{nn}^{(2)}(\omega_{3})=\sum_{\omega_{1},\omega_{2}}[-ie\vec{E}(\omega_{1})\cdot\partial _{\vec{k}}\rho_{nn}^{(1)}(\omega_{2})+ \sum_{m}(\langle n|H_{1}(\omega_{1})|m\rangle\rho_{mn}^{(1)}(\omega_{2})-\rho_{nm}^{(1)}(\omega_{2})\langle m|H_{1}(\omega_{1})|n\rangle)]\frac{\delta(\omega_{3},\omega_{1}+\omega_{2})}{\hbar\omega_{3}},
   \end{aligned}
   \end{equation}
    \end{widetext}
   while the second-order inter-band density matrix reads
   \begin{widetext}
\begin{equation}\label{in2nm}
   \begin{aligned}
     \rho_{nm}^{(2)}(\omega_{3})=\sum_{\omega_{1},\omega_{2}}[ \frac{eE(\omega_{1})}{\hbar\omega_{3}-\epsilon_{nm}}\mathbf{D}_{nm}(\vec{k})\rho_{nm}^{(1)}(\omega_{2})+
     \frac{\langle n|H_{1}(\omega_{1})|m\rangle(\rho_{mm}^{(1)}(\omega_{2})-\rho_{nn}^{(1)}(\omega_{2}))}{\hbar\omega_{3}-\epsilon_{nm}}]
     \delta(\omega_{3},\omega_{1}+\omega_{2}),
   \end{aligned}
   \end{equation}
   \end{widetext}
   respectively. The shift vector $\mathbf{D}_{nm}$ is defined as $\mathbf{D}_{nm}(\vec{k})=-i\partial_{\vec{k}}+\mathbf{a}_{mm}(\vec{k})-\mathbf{a}_{nn}(\vec{k})$ and is invariant under the gauge transformation of Bloch functions. This shift vector characterizes the difference between intracell position matrices within valence and conduction bands. We will ignore the contribution involving three (non-degenerate)bands. The justification for such treatment is that $\langle m|\mathbf{v}|n\rangle\langle n|H_{1}|l\rangle\rho_{lm}^{(1)}$, in which $l\neq m$ and $l\neq n$, does not vanish only when the three atomic orbital hybridize with each other, i.e., three-body interaction which usually is weaker that two-body interaction.

   \section{Injection current}
   The second order charge current is determined by
   \begin{equation}\label{J2}
    \begin{split}
    J^{(2)}_{u}(t)=J_{u,1}(t)+J_{u,2}(t)+J_{u,3}(t)+J_{u,4}(t)= \\
    \frac{e}{\hbar}\partial_{u}\epsilon_{n}(\vec{k})\rho_{nn}^{(2)}(\vec{k},t)
    +\frac{ie}{\hbar}\epsilon_{nm}(\vec{k})A_{nm}^{u}(\vec{k})\rho_{mn}^{(2)}(\vec{k},t) \\
    -\frac{e^{2}E_{v}(t)}{\hbar}f_{nm}^{uv}(\vec{k})\rho_{mn}^{(1)}(\vec{k},t)+\frac{e^{2}E_{v}(t)}{\hbar}g_{n}^{(1)uv}(k,t)
    \end{split}
    \end{equation}
   The current in second line of Eq. 16 is from the second order revision of density matrix under external field, while the third line is resulting from the combination of non-commutativity of position operator and first order revision of density matrix.
   The intra-band current $J_{u,1}$ in frequency domain reads,
   \begin{equation}\label{Injection}
   \begin{aligned}
     J_{u,1}(\omega_{3})=\frac{e}{\hbar}&\sum_{n}\int_{BZ}\frac{d\vec{k}}{(2\pi)^{3}}
     \partial_{u}\epsilon_{n}(\vec{k})\rangle\rho^{(2)}_{nn}(\omega_{3})\\&=J_{intra1}^{(2)}(\omega_{3})+J_{intra2}^{(2)}(\omega_{3})
   \end{aligned}
   \end{equation}
   The first term $J_{intra1}^{(2)}(\omega_{3})$ reads,
   \begin{equation}\label{J11}
   \begin{aligned}
     J_{intra1}^{(2)}(\omega_{3})=\frac{-ie^{2}}{\hbar^{2}\omega_{3}}\delta(\omega_{3},\omega_{1}+\omega_{2})\sum_{\omega_{1},\omega_{2}}\\
     \int_{BZ}\frac{d\vec{k}}{(2\pi)^{3}}
     \partial_{\vec{k}}\epsilon_{n}(\vec{k})\vec{E}(\omega_{1})\cdot\partial_{\vec{k}}\rho^{(1)}_{nn}(\omega_{2}),
   \end{aligned}
   \end{equation}
   The TRIS requires the Hamiltonian and its eigenstates satisfying $h_{0}(\vec{k})$=$h_{0}(-\vec{k})$ and $\epsilon_{n}(\vec{k})$=$\epsilon_{n}(-\vec{k})$, respectively i.e., $\partial_{\vec{k}}\epsilon_{n}(\vec{k})$  ($\partial_{\vec{k}}\rho^{(1)}_{nn}(\vec{k})$) is odd (even) function of momentum $\mathbf{k}$. Therefore, the integrand in Eq.~18 is odd function of momentum $\vec{k}$, and this term is vanishing because of TRIS. The second term $J_{intra2}^{(2)}(\omega_{3})$ in Eq.~17 reads
   \begin{equation}\label{J12}
   \begin{aligned}
     J_{intra2}^{(2)}(\omega_{3})=\sum_{\omega_{1},\omega_{2}}\frac{e}{\hbar^{2}\omega_{3}} \sum_{m}\int_{BZ}\frac{d\vec{k}}{(2\pi)^{3}}
     \partial_{\vec{k}}\epsilon_{n} \\
     [\langle n|H_{1}(\omega_{1})|m\rangle\rho_{mn}^{(1)}(\omega_{2})-c.c.]\delta(\omega_{3}-\omega_{1}-\omega_{2}).
    \end{aligned}
   \end{equation}
    By transformation into time domain, $J_{intra2}^{(2)}(t)$ satisfies equation
    \begin{equation}\label{CPGE}
    \begin{aligned}
      \frac{dJ_{intra2}^{(2)}(t)}{dt}=\sum_{\omega_{3}}(-i\omega_{3})J_{intra2}^{(2)}(\omega_{3})\exp(-i\omega_{3} t)
    \end{aligned}
    \end{equation}
    Since we are interested with dc current, letting $\omega_{3} \rightarrow 0$, the injection current under circularly polarized light $\vec{E}(\omega)=|E|(1,i,0)/\sqrt{2}$ reads
    \begin{widetext}
     \begin{equation}\label{CPGE1}
     \begin{aligned}
       \frac{dJ_{intra2}^{(2)}(t)}{dt}=\frac{ie}{\hbar^{2}}\sum_{\omega_{1},\omega_{2}}\sum_{nm}\int_{BZ}
       \frac{d\vec{k}}{(2\pi)^{3}}\partial_{\mathbf{k}}\epsilon_{nm}(\mathbf{k})
      \rho_{nm}^{(1)}(\omega_{2})\langle m|H_{1}(\omega_{1})|n\rangle]\delta(0,\omega_{1}+\omega_{2}) \\
     =\frac{i\pi e^{3}}{\hbar^{2}}\sum_{mn}\int_{BZ}\frac{d\vec{k}}{(2\pi)^{3}}f_{mn}^{d}(\vec{k})
     \partial_{\vec{k}}\epsilon_{nm}(\vec{k})
      \mathbb{F}_{nm}^{xy}(\vec{k})\vec{E}\times \vec{E}^{*}\delta(\omega-\epsilon_{nm}(\vec{k}))
     \end{aligned}
    \end{equation}
    \end{widetext}
   where $f_{mn}^{d}(\vec{k})=\rho^{(0)}_{mm}(\vec{k})-\rho^{(0)}_{nn}(\vec{k})$. The abelian Berry curvature reads,
   \begin{equation}\label{BC}
   \begin{aligned}
     \mathbb{F}_{nm}^{xy}(\vec{k})=\langle n|x|m\rangle \langle m|y|n\rangle-
     \langle n|y|m\rangle \langle m|x|n\rangle,
   \end{aligned}
   \end{equation}
   which is antisymmetric under exchange of indices \emph{x} and \emph{y}, and it is also odd function of momentum $\vec{k}$ if time-reversals symmetry is preserved. The injection current requires circularly polarized light, and it is vanishing under linearly polarized light. With special condition of constant $f_{mn}(\vec{k})$, the injection current is proportional to the topological charge\cite{CPGE1}. From Eq.~21, it reveals that injection current will be increasing with time. However, injection current can not grow infinitely, because of impurity scattering. The saturated injection current is proportional to the relaxation time.


\section{Shift current}
   The inter-band electronic current contributed by $J_{u,2}=J_{shift}^{(2)}(\omega)+J_{BCD}^{(2)}(\omega)$ in frequency domain reads,
   \begin{widetext}
   \begin{equation}\label{SB}
   \begin{split}
     J_{u,shift}^{(2)}(\omega)=\sum_{\omega_{1},\omega_{2}}\sum_{nm}\int_{BZ}\frac{d\vec{k}}{(2\pi)^{3}}
     \frac{ie}{\hbar}\epsilon_{mn}(\vec{k})A_{nm}^{u}(\vec{k})
     \frac{e\vec{E}(\omega_{1})\cdot\mathbf{D}_{nm}(\vec{k})}{\hbar\omega_{3}-\epsilon_{nm}}\rho_{nm}^{(1)}(\omega_{2}).
   \end{split}
   \end{equation}
  \end{widetext}
  With $\omega=0$,
  Both circularly polarized and linearly polarized optical field can induce shift current, and the second order optical conductivity from shift vector mechanism reads,

   \begin{equation}\label{SB}
   \begin{split}
     \sigma_{uv\lambda}^{(2)}(0)=\frac{\pi e^{3}}{\hbar^{2}}\delta(\omega-\omega_{nm}(\vec{k}))(\rho^{(0)}_{mm}-\rho^{(0)}_{nn})    \\
     [A^{u}_{nm}D_{nm}^{v}A_{nm}^{\lambda}+A^{u}_{mn}D_{mn}^{v}A_{mn}^{\lambda}]
   \end{split}
   \end{equation}
  which is exactly the formula derived by Sipe et al. Both linearly polarized light and circularly polarized light can induce shift current, which is the only mechanism for photocurrent in topologically trivial semiconductor.

  \section{Anomaly current}
   Additionally, we have another term contributing to the inter-band current
  \begin{equation}\label{BCD}
  \begin{split}
    J_{BCD}^{(2)}(\omega_{3})=\frac{ie}{\hbar}\sum_{mn}\sum_{\omega_{1},\omega_{2}}\int_{BZ}\frac{d\textbf{k}}{(2\pi)^{3}}
    \delta(\omega_{3},\omega_{1}+\omega_{2})  \\
   \epsilon_{mn}(\vec{k})A_{mn}^{u}(\vec{k})\frac{\langle n|H_{1}(\omega_{1})|m\rangle[\rho_{mm}^{(1)}(\omega_{2})-\rho_{nn}^{(1)}(\omega_{2})]}{\hbar\omega_{3}-\epsilon_{nm}}
  \end{split}
  \end{equation}
   For the dc current case, $\omega_{3}=0$, the transverse conductivity reads,
   \begin{equation}\label{BCD1}
    \begin{aligned}
     J_{BCD}^{(2)}(0)=\frac{e^{2}}{\hbar}\sum_{\omega_{1},\omega_{2}}\sum_{m}  \\
     \int_{BZ}\frac{d\textbf{k}}{(2\pi)^{3}} \Omega_{mm}\times \vec{E}(\omega_{1})\rho_{mm}^{(1)}(\omega_{2})\delta(0,\omega_{1}+\omega_{2}).
     \end{aligned}
   \end{equation}
   in which $\Omega_{mm}$ is the Berry curvature of band \emph{m}. The term is dubbed as Berry curvature dipole, and the direction $J_{BCD}^{(2)}(0)$ is perpendicular to the direction of optical field. With circularly polarized light $\vec{E}(\omega)=|E|(1,i,0)/\sqrt{2}$, the current along z-direction reads,
    \begin{equation}\label{BCD2}
      J_{z,BCD}^{(2)}(0)=\frac{ie^{3}}{\hbar^{2}\omega}(\partial_{x}\Omega_{mm}^{x}+\partial_{y}\Omega_{mm}^{y})\rho_{mm}^{(0)}
      E(\omega)\times E^{*}(\omega)
    \end{equation}
 For ideal Weyl fermiom, the Berry curvature $\Omega_{mm}^{i}$ is in form of $\frac{k_{i}}{k^{3}}$. In such case, $J_{z,BCD}^{(2)}(0)$ is proportional to 2/3 of the topological charge.
 With circularly polarized light $\mathbf{E}(\omega)=|E|(1,i,0)/\sqrt{2}$, the intra-band from the third line in Eq.~16 reads
\begin{equation}\label{A32}
    \begin{aligned}
     J_{z,BCC}^{(2)}(0)=-\frac{e^{2}E_{v}(t)}{\hbar}f_{nn}^{zv}(\vec{k})\rho_{nn}^{(1)}(\vec{k},t)  \\
     =\frac{ie^{3}}{\hbar^{2}\omega}[\nabla\times\Omega_{mm}]_{z}\rho_{mm}^{(0)}
      E(\omega)\times E^{*}(\omega).
     \end{aligned}
   \end{equation}
which is similar to the photocurrent induced by BCD. However, it characterizes the anisotropy of Berry curvature, which BCD characterizes the gradient of Berry cuvature. Additionally, linearly polarized light can induce shift current only, and it can not induce the injection, $J_{BCD}^{(2)}(0)$, and $J_{BCC}^{(2)}(0)$. All nonlinear Hall current resulting from the gradient and curl of Berry curvature requires circularly polarized light, and has potential application for Terahertz detection because of the pre-factor 1/$\omega$. For materials with TRIS, the total topological charge is zero, the nonlinear Hall current $J_{BCD}^{(2)}(0)$ will be vanishing. However, the $J_{BCC}^{(2)}(0)$ may preserve if the Weyl point is tilted.

 So far, we obtain all known mechanism for photocurrent. However, there are addtionally mechanisms for photocrrent from the non-commutativity of position operator from the third line of Eq.~16.

 \section{3-form tensor}
 Under circularly polarized light $\vec{E}(\omega)=|E|(1,i,0)/\sqrt{2}$, the inter-band photocurrent from the third line in Eq. 16 reads,
\begin{equation}\label{A3}
    \begin{aligned}
     J_{z,nc}^{(2)}(0)=\frac{2i\pi e^{3}}{\hbar^{2}}\sum_{m}\delta(\hbar\omega-\epsilon_{nm}(\vec{k})) \\ \int_{BZ}\frac{d\textbf{k}}{(2\pi)^{3}}f_{mn}^{d}(\vec{k})[f_{mn}^{zx}A_{nm}^{y}-f_{mn}^{zy}A_{nm}^{x}]|E|^{2}.
     \end{aligned}
   \end{equation}
With constant electronic distribution $f_{mn}^{d}(\vec{k})$, the photocurrent from non-commutativity of position operator reads,
 \begin{equation}\label{A31}
    \begin{aligned}
     J_{nc}^{(2)}(0)=\frac{2i\pi e^{3}}{\hbar^{2}}\int_{BZ}\frac{d\textbf{k}}{(2\pi)^{3}}Tr[d\vec{A}\bigwedge \vec{A}]|E|^{2}.
     \end{aligned}
   \end{equation}
Here, $Tr$ labels trace. This current characterizes the integral of 3-form $d\vec{A}\bigwedge \vec{A}$ in the BZ. So far, we can conclude that all gradient, curl and 3-form $d\vec{A}\bigwedge \vec{A}$ can induce NLO. In fact, we can classify them into the same category, i.e., third rank tensor defined with Bloch state (fiber bundle on BZ). Summarily, we developed the theory of NLO in the view of high rank tensor field. However, the detection and application of high rank tensor field are still open questions, and will be left into future works.



\section{Summary}
Motivated by the nonlinear Hall effect induced by Berry curvature dipole (BCD), we studied the photocurrent induced by high rank tensor field in this work. By the quantum kinetic equation, we recovered the shift current and injection current discovered by Sipe et al., and the Berry curvature dipole proposed by Inti Sodemann and Liang Fu. Especially, we further proposed that 3-form tensor like $d\vec{A}\bigwedge \vec{A}$ in which $\vec{A}$ is one-form Berry connection, can also induce photocurrent, in addition to the gradient and curl of Berry curvature. This work will supplement the existing mechanisms for photocurrent. In contrast to the shift current induced by shift vector, all photocurrents induced by gradient/curl of Berry curvature, and high rank tensor $d\vec{A}\bigwedge \vec{A}$ require circularly polarized light and topologically non-trivial band structure, viz. non-vanishing Berry curvature.

\section{Acknowledgements}
This work is the National Natural Science Foundation of China (11604068). T.I. is supported by MEXT via ¡°Exploratory Challenge on Post-K Computer¡± (Frontiers of Basic Science: Challenging the Limits). 




\end{document}